\documentclass{midl}

\usepackage{booktabs}
\usepackage{longtable}
\usepackage{multirow}
\usepackage{float}

\jmlrvolume{}
\jmlryear{2026}
\jmlrpages{}
\jmlrworkshop{}
\editors{}

\title[Conformal Lower Predictive Bounds in Pathology]{A Multi-Cohort Validation of Censoring-Aware Conformal Lower Predictive Bounds for Pathology Survival Models}

\midlauthor{\Name{Mingi Hong}\,\orcid{0009-0009-4121-5765}\Email{me@mghong.dev}\\
\addr Independent Researcher}

\hypersetup{
  pdftitle={A Multi-Cohort Validation of Censoring-Aware Conformal Lower Predictive Bounds for Pathology Survival Models},
  pdfauthor={Mingi Hong},
  pdfsubject={Preprint},
  pdfkeywords={survival analysis, conformal prediction, computational pathology, censoring, multiple-instance learning}
}

\makeatletter
\renewcommand*{\@titlefoot}{\scriptsize
  \textcopyright\ 2026 Mingi Hong. Licensed under
  \href{https://creativecommons.org/licenses/by/4.0/}{CC BY 4.0}.\hfill
  Preprint, July 2026}
\def\ps@jmlrtps{%
  \let\@mkboth\@gobbletwo
  \def\@oddhead{\hfill\scriptsize\textit{Preprint, July 2026}\hfill}%
  \let\@evenhead\@oddhead
  \def\@oddfoot{\@titlefoot}%
  \let\@evenfoot\@oddfoot
}
\makeatother

\begin{document}

\maketitle

\begin{abstract}
Whole-slide survival models commonly provide risk rankings without calibrated statements about individual event times. We evaluate fixed-cutoff drcosarc, a post-hoc conformal wrapper for discrete-time multiple-instance learning survival heads using frozen UNI2-h representations, in an internal 18-configuration sweep across five TCGA cohorts and an external five-configuration evaluation across three CPTAC cohorts. We distinguish configuration--fold--split summaries of the inverse-probability-of-censoring-weighted (IPCW) estimate and median lower predictive bound (LPB) from a hierarchy-aware patient-ensemble estimand of the mean drcosarc--naive LPB difference. At $\alpha=0.1$, the drcosarc IPCW estimate was nearest 0.90 in KIRC, LUAD, and STAD. Patient-ensemble drcosarc--naive intervals excluded zero in KIRC, KIRP, STAD, UCEC, and CPTAC-CCRCC, but included zero in internal LUAD, CPTAC-LUAD, CPTAC-UCEC, and the internal LUSC extension. In a 20-replicate low-censoring semi-synthetic setting with known event times, drcosarc empirical coverage was 0.9129 [0.9053, 0.9207]. An exploratory analysis supported a head-error-by-censoring interaction within that data-generating process. In a two-cohort ABMIL sensitivity analysis, increasing the hazard grid to $K=16$ raised localized marginal IPCW estimates above the prespecified 0.87 threshold and yielded positive paired LPB differences, although worst-group estimates remained below 0.87. Overall, performance was cohort dependent, and its interpretation changed with the patient-level unit, estimand, and censoring assumptions.
\end{abstract}

\section{Introduction}

Pathology survival pipelines increasingly combine frozen foundation-model patch embeddings with multiple-instance learning (MIL) heads \cite{chen2024uni,gustafsson2026benchmarking}. These models are commonly judged by concordance, which measures patient ranking rather than the calibration of a time-valued prediction \cite{harrell1982yield,uno2011cstat,ghawami2026calibration}. A high concordance therefore does not state how long an individual patient will remain event free. Deployment-oriented evaluation requires uncertainty statements about event time in addition to discrimination.

We study a lower predictive bound (LPB) $L(X)$ satisfying the marginal target $P\{T\geq L(X)\}\geq 1-\alpha$, where $T$ is the latent event time. Under right censoring, only $Y=\min(T,C)$ and the event indicator are observed. Finite-sample results for Type-I censoring do not transfer automatically to this general setting. We therefore evaluate an existing imputation-and-weighting construction whose validity is asymptotic and doubly robust under stated regularity and nuisance-consistency conditions \cite{sesia2025doubly}. Real-data coverage is estimated with inverse-probability-of-censoring weighting (IPCW), and semi-synthetic experiments provide complementary evaluation against generated event times.

Prior work addresses conformal inference under several censoring regimes and uncertainty for pathology survival models, but the two strands rarely meet at the level of patient bags. Existing pathology approaches target evidential uncertainty, interval-valued risk outputs, or conformal classification, while imaging-adjacent conformal survival work uses image features within a multimodal covariate vector \cite{xing2026dpsurv,dey2025pathgen,zhang2026truecam,davidov2025general}. This leaves a practical validation gap: censoring-aware latent-time LPBs have not been systematically evaluated with pathology bags as the primary input across MIL heads, cohorts, and external pathology data.

We evaluate the fixed-cutoff drcosarc wrapper around fixed MIL survival heads on frozen UNI2-h bags (Fig.~\ref{fig:pipeline}). The primary questions are whether the resulting LPBs approach the nominal marginal target while remaining informative, and whether they improve patient-level efficiency over naive calibration after accounting for the configuration/fold hierarchy. A four-method comparison spans five internal cohorts and 18 shared-protocol configurations; three CPTAC cohorts assess local recalibration and direct transfer.

Complementary analyses separate distinct sources of uncertainty. Cutoff choice, cross-cohort rank pooling, and risk stratification assess calibration-design choices; a semi-synthetic experiment evaluates known event times while varying censoring and head error; and a two-cohort ABMIL analysis varies the hazard-grid resolution. These analyses do not broaden the validity claim: direct transfer is nonexchangeable, real-data estimates depend on a marginal Kaplan--Meier censoring model, and the semi-synthetic interaction is specific to its data-generating process. Throughout, patients are the exchangeable units, calibration and evaluation roles are patient disjoint, and patient-level ensemble effects preserve the configuration/fold hierarchy.

\section{Related Work}

\paragraph{Conformal inference with censoring.}
Conformal survival methods differ in censoring assumptions and guarantee strength. Cand\`es et al. introduced model-agnostic LPBs with finite-sample marginal coverage under exogenous Type-I censoring \cite{candes2023conformalized}; Gui et al. used covariate-adaptive cutoffs within the same censoring regime \cite{gui2024adaptive}. For general right censoring, Davidov et al. derived finite-sample PAC-type bounds \cite{davidov2025general}, while Sesia and Svetnik combined imputation and weighted conformal calibration to obtain asymptotic double robustness \cite{sesia2025doubly}. Related work studies resampling intervals, distribution recalibration, survival bands, and counterfactual or history-aware targets \cite{qin2025resampling,qi2024csd,sesia2025bands,ren2026counterfactual,wang2026history}. These results are complementary rather than interchangeable. Our implementation uses the fixed-cutoff variant of the Sesia--Svetnik general-right-censoring construction and adopts its asymptotic framework.

\paragraph{Uncertainty in pathology survival models.}
Computational-pathology survival studies mainly evaluate discrimination or calibration of estimated survival probabilities. DPsurv combines UNI2-h representations with prototype-based evidential fusion, but its belief intervals are not conformal latent-time LPBs \cite{xing2026dpsurv}. PathGen uses UNI patch embeddings and synthesized transcriptomic features and reports conformal bounds for risk estimates over four survival intervals \cite{dey2025pathgen}. Its conformal score includes neither a censoring model nor censoring weights and targets a different output from an LPB for the latent event time. TRUECAM combines pathology foundation models with conformal prediction for subtype classification, not survival \cite{zhang2026truecam}. Recent work separately benchmarks pathology foundation models for externally validated survival prediction and examines the calibration of multimodal survival probabilities \cite{gustafsson2026benchmarking,ghawami2026calibration}.

\paragraph{Imaging precedent and remaining validation gap.}
Davidov et al. included one TCGA-BRCA slide per patient, compressed a GigaPath embedding to three principal components, and concatenated it with clinical and genomic variables \cite{davidov2025general}. In that setting, pathology contributes to a 19-variable multimodal vector rather than serving as the primary bag-structured input. The open validation question is how censoring-aware LPBs behave when pathology foundation-model bags are the primary input across MIL heads, cohorts, and external data. We study this question with patient-disjoint calibration, explicit patient-level estimands, and sensitivity analyses for cross-cohort rank pooling, risk-tertile stratification, and survival-head resolution.

\section{Methods}

\begin{figure}[H]
\centering
\includegraphics[width=0.82\linewidth]{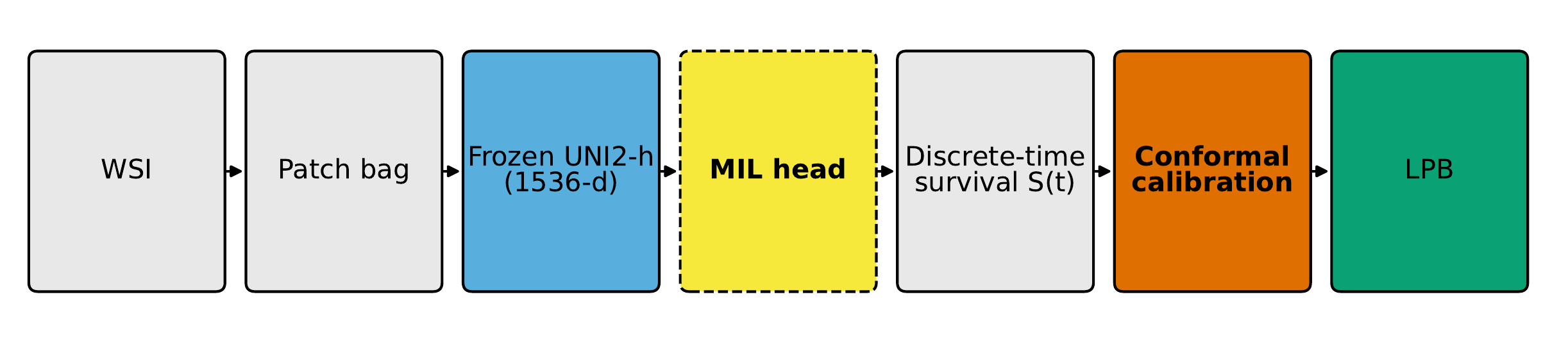}
\caption{Fixed-head validation pipeline. A patient-disjoint calibration set constructs an LPB without updating the encoder or MIL head.}
\label{fig:pipeline}
\end{figure}

\subsection{Prediction setting and target}

For patient $i$, a frozen UNI2-h encoder maps WSI patches to a bag $X_i=\{e_{ij}\}_{j=1}^{n_i}$ with $e_{ij}\in\mathbb{R}^{1536}$. An MIL head trained only on the training split maps this bag to a discrete survival curve on boundaries $0=t_0<t_1<\cdots<t_K$,
\[
\widehat S_i=(1,\widehat S(t_1\mid X_i),\ldots,\widehat S(t_K\mid X_i)).
\]
The primary comparison uses $K=4$ boundaries defined by training-set event-time quartiles. Linear interpolation supplies $\widehat S(t\mid X)$ and the lower time quantile
\[
\widehat q_p(X)=\inf\{t:\widehat S(t\mid X)\leq1-p\};
\]
quantiles that do not cross the grid are truncated at $t_K$. All calibration analyses keep the fitted encoder and head fixed.

All heads optimize discrete-time survival negative log-likelihood. The AEM-regularized survival adaptation \cite{Zhang2025AEM} additionally minimizes
\[
\mathcal{L}_{\mathrm{total}}
=\mathcal{L}_{\mathrm{surv}}+\lambda_t\mathcal{L}_{\mathrm{AEM}},
\qquad
\lambda_t=0.1\,\frac{1+\cos\{\pi t/(T-1)\}}{2},
\]
where $\mathcal{L}_{\mathrm{AEM}}$ is negative attention entropy, $t$ is the zero-indexed epoch, and $T$ is the configured maximum epoch count. This is a survival-task adaptation of AEM rather than a verbatim reproduction of its classification training recipe.

Under right censoring, we observe $Y_i=\min(T_i,C_i)$ and $D_i=\mathbf{1}\{T_i\leq C_i\}$. We target an LPB $L(X)$ at $1-\alpha=0.90$. Because $Y_i\leq T_i$, the directly observable fraction $n^{-1}\sum_i\mathbf{1}\{L(X_i)\leq Y_i\}$ is a lower bound on latent-time coverage. Our primary real-data estimate is
\[
\widehat{\mathrm{Cov}}_{\mathrm{IPCW}}
=\frac{1}{n}\sum_{i=1}^n
\frac{\mathbf{1}\{Y_i\geq L(X_i)\}}
{\widehat G\{L(X_i)^-\}},
\]
where $\widehat G$ is the Kaplan--Meier estimate of the censoring survival function and is floored at $10^{-3}$. This estimator requires an adequate censoring model and may exceed one in small, heavily censored samples. Semi-synthetic experiments instead evaluate $\mathbf{1}\{L(X_i)\leq T_i\}$ using generated event times.

\subsection{Post-hoc censoring-aware calibration}

The wrapper operates on predictions from a fixed head and a disjoint calibration set (Fig.~\ref{fig:pipeline}). We use the fixed-cutoff imputation-and-weighting construction of \citet{sesia2025doubly}, referred to as \emph{drcosarc}, through the released \texttt{candes-fixed} implementation \cite{sesia2025software}. For an observed event, the censoring time is latent; we sample it from the fitted censoring distribution conditional on $C_i>Y_i$. For a censored observation, $C_i=Y_i$ is observed. When $c_0$ is not supplied, the fixed-cutoff path uses the median imputed censoring time. We use that rule and include calibration observations with $C_i\geq c_0$.

For included observations, the nonconformity score and weight are
\[
r_i=\widehat q_\alpha(X_i)-\min(Y_i,c_0),
\qquad w_i=\widehat G(c_0)^{-1}.
\]
In general, the calibration and test weights are $w_i=\widehat G(c_0\mid X_i)^{-1}$ and $w_{\mathrm{new}}=\widehat G(c_0\mid X)^{-1}$; they are constant under our marginal Kaplan--Meier model. The weighted conformal distribution places the test-point mass at $+\infty$:
\[
\widehat P_X
=\sum_{i\in\mathcal I(c_0)}
\frac{w_i}{w_{\mathrm{new}}+\sum_{j\in\mathcal I(c_0)}w_j}\,\delta_{r_i}
+\frac{w_{\mathrm{new}}}{w_{\mathrm{new}}+\sum_{j\in\mathcal I(c_0)}w_j}\,\delta_{+\infty},
\qquad
\widehat q=Q_{1-\alpha}(\widehat P_X),
\]
where $\mathcal I(c_0)=\{i:C_i\geq c_0\}$ and $Q_{1-\alpha}$ is the left $(1-\alpha)$ quantile. Thus, the test weight appears in the denominator for every finite score and enters the numerator only at $+\infty$.
The reported LPB is
\[
L(X)=\min\!\left\{\widehat q_\alpha(X),c_0,
\max\bigl[0,\widehat q_\alpha(X)-\widehat q\bigr]\right\}.
\]
The cap at $\widehat q_\alpha(X)$ supplies the doubly robust adjustment, and the cap at $c_0$ prevents extrapolation beyond the selected censoring horizon.

The cited fixed-cutoff guarantee is asymptotic rather than finite-sample exact. It assumes i.i.d. latent patient triples, conditional independence $T\perp C\mid X$, positivity, and regularity. Its double-robust alternatives require either convergence of the relevant censoring probability, density, and imputation components at the stated rates, or consistency of the conditional event-time $\alpha$-quantile under smoothness conditions. Our marginal Kaplan--Meier estimator does not adjust for covariate-dependent censoring, and selecting $c_0$ from the calibration data does not establish that these conditions hold. We therefore examine empirical sensitivity without interpreting sample splitting as evidence for the theorem's assumptions.

\subsection{Patient bags are the exchangeable units}

The calibration and test units are patients, not patches or slides. Each observation is a triple $(X_i,Y_i,D_i)$. A permutation-invariant pooler makes a prediction insensitive to patch order, but it does not make patches independent observations. Order-sensitive poolers also remain deterministic patient-level predictors when a fixed ordering rule is used. We therefore keep every patient's slides together in training, early validation, conformal calibration, and evaluation splits. This defines the unit of the assumption; it does not make coverage under general right censoring finite-sample exact.

\subsection{Comparators and sensitivity analyses}

Naive calibration, PathGen-C, and drcosarc use the same frozen head. \emph{Naive} treats $Y_i$ as the target, calibrates $r_i=\widehat q_\alpha(X_i)-Y_i$, and returns $\max\{0,\widehat q_\alpha(X)-\widehat q\}$. With the standard split-conformal quantile and exchangeable patients, its marginal coverage for $Y$ transfers conservatively to $T$ because $Y\leq T$; censoring can nevertheless shorten its bounds. \emph{PathGen-C} is the manuscript-specific adaptation of the four-bin conformal risk construction of \citet{dey2025pathgen}: it uses the shared head and maps the conservative end of its interval to an LPB. It is not a native PathGen output. \emph{DPsurv} is trained separately on the same splits and labels and contributes a point-quantile LPB \cite{xing2026dpsurv}. It is an evidential comparator rather than a calibration ablation.

The fixed-cutoff sensitivity sets $c_0$ to the 0.25, 0.50, or 0.75 calibration imputed-censoring quantile; the median is primary. For each cohort, the three split replicates are averaged within each of the 18 configurations and five folds. A nonparametric bootstrap then resamples the resulting 90 configuration--fold units per cohort to estimate paired 0.25-minus-0.50 and 0.75-minus-0.50 contrasts in the IPCW estimate and median LPB. These contrasts characterize sensitivity to the cutoff choice and are not a prespecified robustness criterion. A semi-synthetic Weibull mechanism probes covariate-dependent censoring, but the real-data primary analysis uses marginal Kaplan--Meier censoring survival. Cross-cohort rank pooling is evaluated empirically by paired pooled-minus-per-cohort contrasts with hierarchy-aware intervals and a prespecified 0.87 floor for the worst cohort; this analysis does not establish conditional validity.

The exploratory patient-grouped risk-tertile analysis at $\alpha=0.1$ spans all five primary cohorts, 18 configurations, five folds, and three calibration/evaluation splits. Within each split, tertile cutpoints are learned only from predicted risk in the calibration set; evaluation outcomes are not used to define the groups. For evaluation cell $g$, we compute the bounded H\'ajek estimator
\[
\widehat{\mathrm{Cov}}_{g,\mathrm{Hajek}}
=\frac{\sum_{i\in g} w_i\mathbf{1}\{Y_i\geq L(X_i)\}}
{\sum_{i\in g} w_i},
\qquad
w_i=\widehat G\{L(X_i)^-\}^{-1}.
\]
Available split ratios are averaged within configuration--fold and then equally across 90 units per cohort--tertile. A 10{,}000-replicate configuration/fold bootstrap with seed 20260723 gives cellwise intervals and a maximum-deviation simultaneous upper bound across 15 cells; empty-cell handling appears in the supplement. These are empirical diagnostics, not conditional coverage guarantees.

The head-resolution sensitivity retrains ABMIL with $K\in\{4,8,16\}$ on KIRC and LUAD across five patient-grouped outer folds; each outer-fold fit uses a single initialization seed. Here, localized calibration selects an adaptive probability level within each patient group, rather than applying the standard fixed-cutoff drcosarc construction globally. The $K=4$ setting provides the baseline grid resolution. For $K=8$ and $K=16$, the prespecified marginal criteria require a patient-grouped localized IPCW estimate of at least 0.87 and a positive $\Delta\mathrm{LPB}=\mathrm{LPB}_{\mathrm{localized}}-\mathrm{LPB}_{\mathrm{drcosarc}}$ at the same $K$, with an interval excluding zero. The prespecified conditional criterion additionally requires a worst-group estimate of at least 0.87.

\subsection{Evaluation endpoints}

Coverage is paired with efficiency because a trivially small LPB is uninformative. We compute one summary for each combination of configuration, outer fold, and calibration/evaluation split replicate. This configuration--fold--split summary contains the IPCW estimate across evaluation patients and the median patient LPB. Table~\ref{tab:master-comparison} averages the IPCW estimate and median LPB separately across 18 configurations $\times$ 5 folds $\times$ 3 split replicates, or 270 summaries per cohort and method. Table~\ref{tab:paired-efficiency} targets a distinct patient-ensemble estimand: split-replicate LPBs are first combined within each configuration/fold for each patient, after which the all-patient mean drcosarc--naive LPB difference and hierarchy-aware uncertainty are estimated. Harrell and Uno C-indices measure discrimination \cite{harrell1982yield,uno2011cstat}; IBS, integrated binomial log-likelihood, and D-calibration characterize the fitted survival head \cite{graf1999assessment,kvamme2021continuous,haider2020effective}. Semi-synthetic coverage is evaluated directly against generated $T$; interaction intervals use replicate-cluster bootstrap samples.

\section{Experiments}

\subsection{Data and protocol}

The primary internal comparison uses TCGA KIRC, KIRP, LUAD, STAD, and UCEC with curated TCGA-CDR overall-survival labels \cite{liu2018tcgacdr}; LUSC contributes only an internal extension to the paired-efficiency analysis. All experiments use frozen 1536-dimensional UNI2-h embeddings \cite{chen2024uni,mahmoodlab2025uni2}. The outer five-fold split is patient grouped and event stratified. No site field was available, so site stratification was not possible; unmodeled site signatures can affect pathology-model accuracy and bias \cite{howard2021site}. The allocation is approximately 70\% model fitting, 10\% patient-grouped early validation, and 20\% outer held-out data; early validation is not event stratified. Each held-out prediction set is divided into patient-disjoint, event-stratified calibration and evaluation subsets for three split replicates, yielding approximately 70/10/10/10 after rounding.

The internal sweep comprises 18 configurations $\times$ five primary cohorts $\times$ five folds, for 450 trained heads. Each head has a single training run, initialized with seed 42. Every configuration uses a 512-dimensional adaptor, a linear $K=4$ hazard head, batch size 1, Adam at $10^{-4}$, discrete-time negative log-likelihood, and the same early-stopping rule \cite{kvamme2021continuous}; the AEM survival adaptation additionally uses the entropy regularizer defined above. Early stopping uses validation Harrell C-index between epochs 20 and 100. External validation fixes the TCGA heads and bins and evaluates a five-configuration subset on CPTAC-CCRCC, CPTAC-LUAD, and CPTAC-UCEC. Variation across outer folds therefore reflects changes in patient data partitions, not repeated optimization-seed variation; the latter was not evaluated.

The 120 cohort--fold--split-replicate partitions have no patient-ID overlap among training, early validation, calibration, and evaluation roles within a partition. The primary target is 0.90 coverage. Real-data coverage is estimated by IPCW, whereas semi-synthetic coverage is evaluated directly against generated event times.

\subsection{Primary comparison}

The primary question is whether fixed-cutoff calibration yields informative LPBs with an IPCW estimate near 0.90. Table~\ref{tab:master-comparison} averages the IPCW estimate and median LPB separately across 270 configuration--fold--split summaries per cohort and method: 18 configurations $\times$ five folds $\times$ three calibration/evaluation split replicates. These averages do not describe one pooled patient distribution. The drcosarc estimate was nearest 0.90 in KIRC, LUAD, and STAD, whereas every method produced estimates above 0.90 in the heavily censored KIRP and UCEC cohorts. Thus, proximity to the target was cohort dependent under the specified censoring model.

\begin{table}[H]
\centering
\caption{Primary internal comparison at $\alpha=0.1$. Entries are mean IPCW estimate / mean median LPB in days across configuration--fold--split summaries. PathGen-C is the manuscript-specific clamped one-sided adaptation; DPsurv is trained separately.}
\label{tab:master-comparison}
\small
\begin{tabular}{lcccc}
\toprule
Cohort & Naive & PathGen-C & DPsurv point & drcosarc \\
\midrule
KIRC & 0.949 / 135 & 0.977 / 100 & 0.955 / 215 & 0.913 / 313 \\
KIRP & 0.995 / 54 & 0.996 / 87 & 0.979 / 270 & 0.985 / 253 \\
LUAD & 0.944 / 39 & 0.982 / 5 & 0.948 / 174 & 0.922 / 109 \\
STAD & 0.959 / 43 & 0.998 / 8 & 0.945 / 97 & 0.933 / 89 \\
UCEC & 0.984 / 140 & 0.986 / 180 & 1.001 / 270 & 0.982 / 286 \\
\bottomrule
\end{tabular}
\end{table}

Naive, PathGen-C, and drcosarc share the fitted survival head, so their differences isolate LPB construction. PathGen-C is the manuscript-specific one-sided adaptation rather than a native PathGen output. DPsurv instead uses a separately trained evidential head and serves as a contextual comparator.

\subsection{Patient-level paired efficiency}

The second question is whether drcosarc improves patient-level efficiency over naive calibration. Table~\ref{tab:paired-efficiency} uses a different estimand from Table~\ref{tab:master-comparison}: it combines split replicates within configuration/fold, forms patient-ensemble LPBs, and estimates the all-patient mean drcosarc--naive difference with hierarchy-aware uncertainty. Intervals excluded zero in KIRC, KIRP, STAD, UCEC, and CPTAC-CCRCC. They included zero in internal LUAD, CPTAC-LUAD, CPTAC-UCEC, and the internal LUSC extension (Fig.~\ref{fig:main-results}c). The supported efficiency gains therefore do not extend uniformly across cohorts.

\begin{table}[H]
\centering
\caption{Patient-ensemble paired efficiency of drcosarc versus naive calibration. $\Delta$ is the all-patient mean LPB difference; intervals preserve the configuration/fold hierarchy.}
\label{tab:paired-efficiency}
\small
\begin{tabular}{llrc}
\toprule
Scope & Cohort & Patients & $\Delta$ LPB, days [95\% CI] \\
\midrule
Internal & KIRC & 509 & $+173.58$ [$+137.08$, $+211.90$] \\
Internal & KIRP & 270 & $+139.76$ [$+111.67$, $+169.99$] \\
Internal & LUAD & 456 & $+7.35$ [$-24.32$, $+35.17$] \\
Internal & STAD & 358 & $+24.90$ [$+7.94$, $+40.27$] \\
Internal & UCEC & 502 & $+157.41$ [$+117.93$, $+197.76$] \\
External local & CCRCC & 198 & $+301.53$ [$+226.84$, $+378.18$] \\
External local & LUAD & 199 & $+40.40$ [$-0.44$, $+109.67$] \\
External local & UCEC & 93 & $+14.07$ [$-6.43$, $+82.16$] \\
Internal extension & LUSC & 463 & $+2.74$ [$-5.87$, $+49.78$] \\
\bottomrule
\end{tabular}
\end{table}

\subsection{External validation}\label{sec:external-validation}

External validation asks how local recalibration and direct transfer differ under site shift. With local CPTAC calibration, the drcosarc mean IPCW estimate / mean median LPB across configuration--fold--split summaries was 0.954 / 322 days in CPTAC-CCRCC, 1.012 / 120 in CPTAC-LUAD, and 1.014 / 275 in CPTAC-UCEC. The corresponding naive values were 0.970 / 41, 0.981 / 40, and 0.972 / 184. Under direct TCGA-to-CPTAC transfer, drcosarc estimates were 0.960 / 330, 1.002 / 54, and 1.009 / 226, respectively. Direct transfer makes calibration and evaluation distributions nonexchangeable by design; these IPCW estimates therefore measure empirical performance under the observed shift rather than guarantee coverage. CPTAC-UCEC has only 10 events, so its estimates remain illustrative. Complete values appear in Supplementary Table~\ref{tab:supp-external}.

\subsection{Semi-synthetic controls}\label{sec:interaction-law}

\begin{figure}[t]
\centering
\begin{minipage}[t]{0.31\linewidth}
\centering
\includegraphics[width=\linewidth]{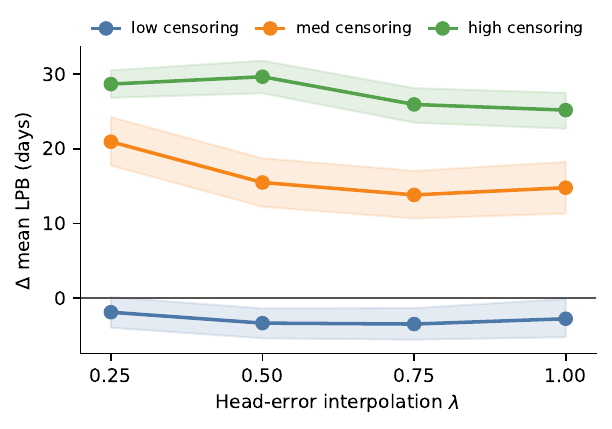}\\[-1mm]
\textbf{(a)} Head error $\times$ censoring
\end{minipage}\hfill
\begin{minipage}[t]{0.31\linewidth}
\centering
\includegraphics[width=\linewidth]{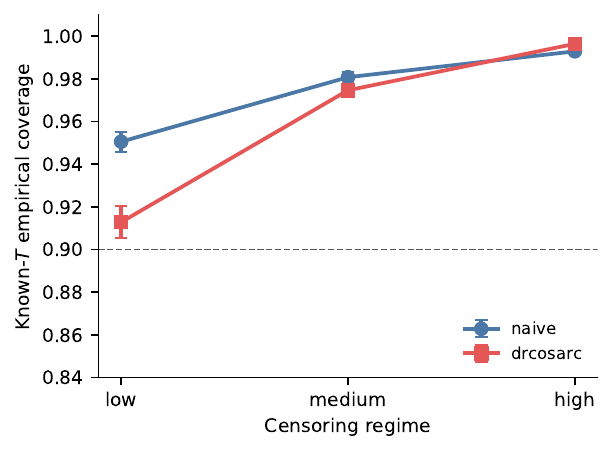}\\[-1mm]
\textbf{(b)} Known-$T$ coverage
\end{minipage}\hfill
\begin{minipage}[t]{0.31\linewidth}
\centering
\includegraphics[width=\linewidth]{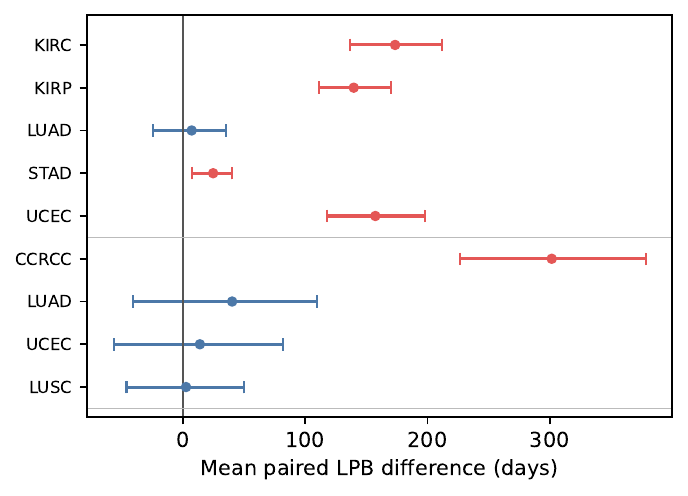}\\[-1mm]
\textbf{(c)} Patient-ensemble paired effects
\end{minipage}
\caption{Semi-synthetic controls and patient-level paired efficiency at $\alpha=0.1$. (a) Mean drcosarc--naive LPB differences across head-error interpolation levels and censoring regimes; bands show uncertainty across replicates. (b) Empirical coverage against generated event times; the dashed line marks 0.90. (c) Mean patient-ensemble drcosarc--naive LPB differences with hierarchy-aware intervals; red intervals exclude zero and blue intervals include zero. Horizontal dividers separate primary internal, external local, and LUSC extension results.}
\label{fig:main-results}
\end{figure}

The semi-synthetic experiment evaluates coverage against known event times within a controlled data-generating process (DGP). In the baseline low-censoring, $\alpha=0.1$ cell, known-$T$ empirical coverage across 20 replicates was 0.9506 for naive calibration and 0.9129 for drcosarc; the drcosarc replicate-bootstrap interval was [0.9053, 0.9207] (Fig.~\ref{fig:main-results}b).

An exploratory head-error-by-censoring model used replicate-cluster bootstrap samples. The interaction coefficient for the mean-LPB drcosarc--naive contrast was 7.43 [6.33, 8.42] days at $\alpha=0.1$ and 15.10 [13.44, 16.78] days at $\alpha=0.2$ (Fig.~\ref{fig:main-results}a). These coefficients support an interaction within this DGP; they do not establish the same relationship in TCGA or CPTAC.

\subsection{Cutoff and calibration-design sensitivity}

The cutoff sensitivity evaluates all primary cohorts, configurations, folds, split replicates, and both alpha levels at imputed-censoring quantiles 0.25, 0.50, and 0.75; the median cutoff is primary. Paired contrasts subtract the 0.50 result from the 0.25 or 0.75 result after averaging split replicates within each configuration/fold. At $\alpha=0.1$, the largest absolute IPCW-estimate contrast was 0.025 (KIRC, $q=0.25$); at $\alpha=0.2$, it was 0.076 (KIRP, $q=0.75$). LPB shifts were materially larger and cohort dependent, so the analysis does not establish robustness to cutoff choice. A Weibull covariate-dependent-censoring sensitivity was evaluated only in the semi-synthetic setting; the real-data estimates continue to rely on the marginal Kaplan--Meier censoring model.

\begin{figure}[H]
\centering
\includegraphics[width=0.38\linewidth]{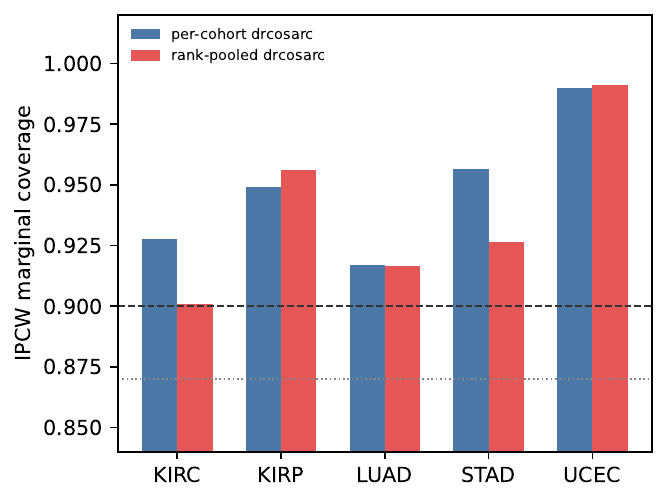}
\caption{Per-cohort versus rank-pooled drcosarc IPCW estimates at $\alpha=0.1$; lines mark 0.90 (dashed) and 0.87 (dotted). Hierarchy-aware contrasts appear in text.}
\label{fig:rank-pooling}
\end{figure}

As a descriptive exploratory analysis, cross-cohort rank pooling reduced the drcosarc IPCW estimate weighted across configuration--fold--split summaries from 0.948 to 0.938, using 270 summaries per cohort and method. The worst pooled cohort was KIRC at 0.901, above the prespecified 0.87 empirical threshold (Fig.~\ref{fig:rank-pooling}). Hierarchy-aware pooled-minus-per-cohort IPCW-estimate contrasts were KIRC $-0.025$ [$-0.033$, $-0.017$], KIRP $+0.007$ [$+0.001$, $+0.013$], LUAD $0.000$ [$-0.007$, $+0.007$], STAD $-0.030$ [$-0.039$, $-0.022$], and UCEC $0.000$ [$-0.006$, $+0.006$]. The negative intervals for KIRC and STAD indicate lower estimates under pooling in those cohorts, but the analysis provides no distribution-free conditional guarantee.

At $\alpha=0.1$, nested-bootstrap intervals and simultaneous upper bounds placed all 15 localized H\'ajek risk-tertile estimates below 0.90. Estimates ranged from 0.251 [0.216, 0.288] for KIRP low risk to 0.800 [0.766, 0.832] for KIRC high risk; the largest simultaneous 95\% upper bound was 0.889. Low risk was lowest in every cohort. This is joint empirical undercoverage under the specified hierarchy, not distribution-free conditional validity.

\subsection{Two-cohort head-resolution sensitivity}

This sensitivity analysis is limited to ABMIL on KIRC and LUAD, with $K\in\{4,8,16\}$ and a single initialization seed for each outer-fold fit. At $K=4$, the localized marginal IPCW estimate was below 0.87 in KIRC (0.723 [0.610, 0.855]) and LUAD (0.794 [0.744, 0.845]). At $K=8$, the estimate remained below 0.87 in KIRC (0.785 [0.642, 0.920]) and LUAD (0.842 [0.779, 0.908]), despite positive $\Delta\mathrm{LPB}$ values relative to standard drcosarc at the same $K$. At $K=16$, both cohorts met the prespecified marginal point-estimate and paired-efficiency criteria: estimates were 0.897 [0.863, 0.932] in KIRC and 0.891 [0.856, 0.926] in LUAD, with localized-minus-drcosarc differences of $+254$ [$+48$, $+516$] and $+91$ [$+41$, $+135$] days, respectively. Worst-group estimates remained below 0.87 in both cohorts (0.740 and 0.766), so the sensitivity analysis does not support conditional-validity recovery (Supplementary Tables~\ref{tab:supp-resolution}--\ref{tab:supp-resolution-head}).

Increasing $K$ from 4 to 16 slightly reduced C-index while improving IBS in both cohorts. The $K=16$ heads contained no zero-event bins, with mean events per bin of 7.58 in KIRC and 8.69 in LUAD. The change therefore cannot be attributed to uniformly better discrimination. Because the reported IPCW estimate can exceed one, it is not raw head-quantile coverage and does not show that quantile overoptimism decreases with $K$.

\subsection{Configuration breadth}

The 18 configurations span attention, graph/spatial, transformer, and state-space families (Supplementary Table~\ref{tab:supp-configurations}). All use the same adaptor, $K=4$ hazard head, survival loss, and optimization protocol; the AEM adaptation adds its scheduled entropy regularizer. The sweep evaluates post-hoc compatibility under a shared survival protocol rather than each architecture's native training recipe.

\section{Discussion}

We evaluated fixed-cutoff conformal LPBs for general right censoring in pathology MIL. Because naive calibration, PathGen-C, and drcosarc share a fitted head, their differences reflect LPB construction rather than discrimination; DPsurv provides context from a separately trained head. Under the configuration--fold--split-summary estimand, the drcosarc IPCW estimate was nearest 0.90 in KIRC, LUAD, and STAD, whereas all methods produced estimates above 0.90 in KIRP and UCEC. The patient-ensemble analysis addressed a different estimand: LPB-gain intervals included zero in internal LUAD, CPTAC-LUAD, CPTAC-UCEC, and internal LUSC, showing that supported efficiency gains were not uniform.

The semi-synthetic experiment supplies the most direct coverage evidence because the event times are known. In its low-censoring baseline, drcosarc attained 0.9129 coverage at $\alpha=0.1$, and replicate-clustered analysis supported a positive head-error-by-censoring interaction. That mechanism evidence is confined to the simulated data-generating process; real-cohort IPCW estimates remain sensitive to the fixed cutoff and censoring model.

The subgroup and resolution analyses narrow the interpretation of favorable marginal results. At $K=16$, the two-cohort ABMIL probe met its marginal IPCW-estimate and paired-efficiency criteria despite slightly lower C-index, but its worst-group estimates remained below 0.87. Likewise, all simultaneous upper bounds in the five-cohort H\'ajek risk-tertile analysis were below 0.90, with the low-risk groups most vulnerable. Finer discretization therefore did not establish conditional validity, and sparse tertile-specific events and unstable censoring weights remain important limitations.

External and pooled analyses further separate empirical transport from conformal validity. Local CPTAC recalibration preserves within-cohort exchangeability subject to the censoring model, whereas direct TCGA-to-CPTAC transfer is nonexchangeable by design and CPTAC-UCEC has few events. Rank pooling cleared the 0.87 worst-cohort floor, but hierarchy-aware intervals showed reductions in KIRC and STAD. Together with single-seed optimization, a mostly shared rather than architecture-native training protocol, censoring-model dependence, and the absence of prospective validation, these findings support estimand-specific patient-level validation rather than a universal coverage claim.

\section{Conclusion}

Fixed-cutoff censoring-aware calibration can provide informative post-hoc LPBs for pathology MIL survival heads without changing their discrimination, although the coverage--efficiency trade-off varies across cohorts. Hierarchy-aware patient-ensemble gains were supported in five cohorts; the two-cohort ABMIL sensitivity retained worst-group undercoverage despite improved marginal behavior. More broadly, validating conformal survival predictions requires treating patients as the statistical units, stating the censoring assumptions, and matching each conclusion to its estimand.

\clearpage
\section*{Declarations}

\paragraph{Competing interests.}
The author declares no competing interests.

\paragraph{Ethics statement.}
This study used publicly available, de-identified data and precomputed pathology representations. No new participant recruitment or data collection was performed.

\paragraph{Data availability.}
The clinical and outcome data analyzed in this study are publicly available from The Cancer Genome Atlas (TCGA) and the Clinical Proteomic Tumor Analysis Consortium (CPTAC), subject to their respective access terms. Pre-extracted UNI2-h representations for TCGA and CPTAC were obtained from the MahmoodLab UNI2-h features repository on Hugging Face and remain subject to that repository's access and license terms \citep{mahmoodlab2025uni2features}.

\paragraph{Code availability.}
The analysis code will be made publicly available in a versioned repository upon publication.

\paragraph{Author contributions.}
Mingi Hong conceived the study, developed the methodology and software, conducted the analyses and validation, curated the data, created the visualizations, and wrote and revised the manuscript.

\paragraph{License.}
\textcopyright{} 2026 Mingi Hong. This preprint is licensed under the \href{https://creativecommons.org/licenses/by/4.0/}{Creative Commons Attribution 4.0 International License}.

\bibliography{references}

\clearpage
\appendix
\section{Supplementary Material}
\setcounter{table}{0}
\renewcommand{\thetable}{S\arabic{table}}
\renewcommand{\theHtable}{S\arabic{table}}
\setcounter{figure}{0}
\renewcommand{\thefigure}{S\arabic{figure}}
\renewcommand{\theHfigure}{S\arabic{figure}}

\subsection{Reproducibility protocol}

The internal sweep contains 18 configurations, five primary TCGA cohorts, and five patient-grouped, event-stratified outer folds. No site variable was available. Every configuration--cohort--outer-fold fit uses frozen 1536-dimensional UNI2-h features, a 512-dimensional adaptor, a linear $K=4$ hazard head, batch size 1, Adam at $10^{-4}$, discrete-time negative log-likelihood, and early stopping between epochs 20 and 100. The AEM-regularized survival adaptation additionally uses negative attention entropy with a weight that cosine-anneals from 0.1 to zero. Each fit has a single training run, initialized with seed 42. The allocation is approximately 70\% model fitting, 10\% patient-grouped early validation, and 20\% outer held-out data. Early validation is not event stratified. Each held-out prediction set is split three times into patient-disjoint, event-stratified calibration and evaluation subsets.

Within each of the 120 cohort--fold--split-replicate partitions, every patient is assigned to only one of the training, early-validation, calibration, or evaluation roles. Patients may appear in different roles across outer folds, as required by cross-validation.

With marginal Kaplan--Meier censoring survival $\widehat G$, the calibration and test weights equal $\widehat G(c_0)^{-1}$. Every conformal run attaches the test-point mass to $+\infty$:
\[
\widehat P_X
=\sum_{i\in\mathcal I(c_0)}
\frac{w_i}{w_{\mathrm{new}}+\sum_{j\in\mathcal I(c_0)}w_j}\,\delta_{r_i}
+\frac{w_{\mathrm{new}}}{w_{\mathrm{new}}+\sum_{j\in\mathcal I(c_0)}w_j}\,\delta_{+\infty}.
\]
\subsection{Comparator definitions}

PathGen-C is the manuscript-specific, shared-head, four-bin, non-censor-aware one-sided adaptation. The primary comparison uses the clamped variant. Supplementary comparisons also consider an unclamped implementation, native two-sided PathGen/MCAT coverage, and a derived one-sided LPB; none is a native PathGen/MCAT LPB output.

DPsurv uses a separately trained evidential head under the shared patient-grouped splits and survival labels.

\subsection{Head and external results}

\begin{table}[H]
\centering
\caption{Primary head metrics, reported as Harrell C-index / IBS. Shared denotes the head used by naive calibration, PathGen-C, and drcosarc; DPsurv is trained separately.}
\label{tab:supp-primary-details}
\small
\begin{tabular}{lcc}
\toprule
Cohort & Shared & DPsurv \\
\midrule
KIRC & 0.70 / 0.17 & 0.68 / 0.19 \\
KIRP & 0.77 / 0.11 & 0.76 / 0.15 \\
LUAD & 0.57 / 0.23 & 0.53 / 0.21 \\
STAD & 0.55 / 0.21 & 0.51 / 0.21 \\
UCEC & 0.68 / 0.10 & 0.69 / 0.15 \\
\bottomrule
\end{tabular}
\end{table}

\begin{table}[H]
\centering
\caption{CPTAC results at $\alpha=0.1$. Entries are mean IPCW estimate / mean median LPB in days across configuration--fold--split summaries. Local calibration and direct transfer are distinct designs.}
\label{tab:supp-external}
\small
\begin{tabular}{llcc}
\toprule
Cohort & Design & Naive & drcosarc \\
\midrule
CCRCC & Local & 0.970 / 41 & 0.954 / 322 \\
CCRCC & Transfer; no recalibration & 0.976 / 101 & 0.960 / 330 \\
LUAD & Local & 0.981 / 40 & 1.012 / 120 \\
LUAD & Transfer; no recalibration & 0.987 / 6 & 1.002 / 54 \\
UCEC & Local & 0.972 / 184 & 1.014 / 275 \\
UCEC & Transfer; no recalibration & 0.995 / 77 & 1.009 / 226 \\
\bottomrule
\end{tabular}
\end{table}

IPCW values above one reflect estimator variability under heavy censoring. Transfer separates calibration and evaluation distributions and is an empirical site-shift stress test rather than a theoretical validation.

\subsection{Semi-synthetic and cutoff controls}

The semi-synthetic experiment uses replicate-level DGP settings, split sizes, random-number seeds, observed censoring, known-$T$ coverage, LPB summaries, and selected cutoff values. Head-noise and conformal-imputation random streams are separated. The exploratory interaction estimates use replicate-cluster bootstrap samples.

The cutoff sensitivity evaluates imputed-censoring quantiles 0.25, 0.50, and 0.75 for every primary cohort, configuration, fold, split replicate, and alpha level. The median is primary. Adaptive risk-tertile outputs are separately labelled exploratory and are not interchangeable with the fixed-cutoff primary method.

Table~\ref{tab:supp-cutoff-sensitivity} reports paired contrasts at $\alpha=0.1$, defined as the indicated $q$ minus $q=0.50$. After averaging the three splits within each configuration/fold, a 10{,}000-replicate paired nonparametric bootstrap with base seed 20260723 resampled the resulting 90 configuration--fold units per cohort to form 95\% confidence intervals. No multiplicity adjustment was applied, and the intervals are not coverage guarantees.

\begin{table}[H]
\centering
\caption{Fixed-cutoff sensitivity at $\alpha=0.1$. Contrasts are $q$ minus $q=0.50$; brackets give 95\% nonparametric bootstrap confidence intervals.}
\label{tab:supp-cutoff-sensitivity}
\small
\setlength{\tabcolsep}{3pt}
\begin{tabular}{lc@{\hspace{8pt}}cc}
\toprule
Cohort & $q$ & \shortstack{$\Delta$ coverage\\[1pt] [95\% bootstrap CI]} & \shortstack{$\Delta$ median LPB, days\\[1pt] [95\% bootstrap CI]} \\
\midrule
KIRC & 0.25 & $+0.025$ [$+0.015$, $+0.034$] & $-53$ [$-85$, $-23$] \\
KIRC & 0.75 & $-0.003$ [$-0.017$, $+0.010$] & $+30$ [$-29$, $+87$] \\
KIRP & 0.25 & $+0.023$ [$+0.008$, $+0.039$] & $-111$ [$-137$, $-86$] \\
KIRP & 0.75 & $+0.023$ [$+0.006$, $+0.041$] & $-172$ [$-229$, $-118$] \\
LUAD & 0.25 & $+0.019$ [$+0.012$, $+0.027$] & $-38$ [$-55$, $-21$] \\
LUAD & 0.75 & $+0.017$ [$+0.004$, $+0.031$] & $-42$ [$-67$, $-18$] \\
STAD & 0.25 & $+0.004$ [$-0.001$, $+0.010$] & $+3$ [$-4$, $+9$] \\
STAD & 0.75 & $+0.002$ [$-0.005$, $+0.010$] & $-2$ [$-14$, $+12$] \\
UCEC & 0.25 & $-0.007$ [$-0.019$, $+0.005$] & $-60$ [$-88$, $-32$] \\
UCEC & 0.75 & $+0.024$ [$+0.015$, $+0.033$] & $+45$ [$+6$, $+87$] \\
\bottomrule
\end{tabular}
\end{table}

\begin{table}[H]
\centering
\caption{Patient-grouped localized head-resolution sensitivity. Entries are marginal IPCW estimates [95\% intervals]. $\Delta\mathrm{LPB}=\mathrm{LPB}_{\mathrm{localized}}-\mathrm{LPB}_{\mathrm{drcosarc}}$ at the same $K$. Dashes indicate quantities unavailable for the $K=4$ baseline.}
\label{tab:supp-resolution}
\small
\begin{tabular}{llccc}
\toprule
$K$ & Cohort & Marginal IPCW estimate & Worst group & $\Delta$ LPB, days [95\% CI] \\
\midrule
4 & KIRC & 0.723 [0.610, 0.855] & -- & -- \\
4 & LUAD & 0.794 [0.744, 0.845] & -- & -- \\
8 & KIRC & 0.785 [0.642, 0.920] & 0.604 & $+542$ [$+127$, $+1034$] \\
8 & LUAD & 0.842 [0.779, 0.908] & 0.645 & $+237$ [$+102$, $+402$] \\
16 & KIRC & 0.897 [0.863, 0.932] & 0.740 & $+254$ [$+48$, $+516$] \\
16 & LUAD & 0.891 [0.856, 0.926] & 0.766 & $+91$ [$+41$, $+135$] \\
\bottomrule
\end{tabular}
\end{table}

\begin{table}[H]
\centering
\caption{Head quality and event budget for the endpoint resolutions in the probe.}
\label{tab:supp-resolution-head}
\small
\begin{tabular}{llccccc}
\toprule
$K$ & Cohort & C-index & IBS & Mean events/bin & Min events/bin & Zero-event bins \\
\midrule
4 & KIRC & 0.699 & 0.221 & 30.3 & 30.0 & 0 \\
4 & LUAD & 0.586 & 0.238 & 34.8 & 32.0 & 0 \\
16 & KIRC & 0.674 & 0.207 & 7.58 & 7.0 & 0 \\
16 & LUAD & 0.573 & 0.210 & 8.69 & 3.0 & 0 \\
\bottomrule
\end{tabular}
\end{table}

The IPCW implementation averages $\mathbf{1}\{Y\geq L(X)\}/\widehat G\{L(X)^-\}$ and can exceed one. We therefore do not interpret these outputs as raw head-quantile coverage and do not infer that increasing $K$ reduces raw $\alpha$-quantile overoptimism.

\subsection{Exploratory calibration-design analyses}

Figure~\ref{fig:rank-pooling} presents descriptive cross-cohort rank pooling. The rank-pooled analysis comprises 18 configurations, five folds, three patient-grouped calibration splits, and five cohorts, yielding 270 configuration--fold--split summaries per cohort and method. Its overall summary-weighted IPCW estimate was 0.938 versus 0.948 with per-cohort calibration, and its worst pooled cohort was KIRC at 0.901, above 0.87. Table~\ref{tab:supp-rank-pooling} shows that pooled-minus-per-cohort IPCW-estimate contrasts were negative for KIRC and STAD. After averaging the three splits within each configuration/fold, a paired nonparametric bootstrap resampled the resulting 90 configuration--fold units per cohort to form 95\% confidence intervals. No multiplicity adjustment was applied. The intervals reflect the analysis hierarchy, but neither they nor the worst-cohort threshold provide a distribution-free conditional guarantee.

\begin{table}[H]
\centering
\caption{Cross-cohort rank-pooling sensitivity at $\alpha=0.1$. Entries are pooled minus per-cohort contrasts with 95\% nonparametric bootstrap confidence intervals.}
\label{tab:supp-rank-pooling}
\small
\setlength{\tabcolsep}{3pt}
\begin{tabular}{lcc}
\toprule
Cohort & \shortstack{$\Delta$ coverage\\[1pt] [95\% interval]} & \shortstack{$\Delta$ median LPB, days\\[1pt] [95\% interval]} \\
\midrule
KIRC & $-0.025$ [$-0.033$, $-0.017$] & $+125$ [$+88$, $+165$] \\
KIRP & $+0.007$ [$+0.001$, $+0.013$] & $-9$ [$-17$, $-1$] \\
LUAD & $0.000$ [$-0.007$, $+0.007$] & $-1$ [$-12$, $+10$] \\
STAD & $-0.030$ [$-0.039$, $-0.022$] & $+40$ [$+28$, $+53$] \\
UCEC & $0.000$ [$-0.006$, $+0.006$] & $+38$ [$+19$, $+59$] \\
\bottomrule
\end{tabular}
\end{table}

The patient-grouped risk-tertile analysis at $\alpha=0.1$ covered five cohorts, 18 configurations, five folds, and three splits. Tertile cutpoints were learned only from calibration-set predicted risk, without evaluation outcomes. Each H\'ajek ratio was first computed within a configuration--fold--split cell. Available ratios were averaged within each configuration--fold, followed by an equal average of the 90 configuration--fold values for each cohort--tertile combination. A nested 10{,}000-replicate bootstrap with seed 20260723 resampled configurations and folds to form cellwise percentile intervals. The simultaneous upper bounds add the 95th percentile of the maximum centered deviation across the 15 cohort--tertile cells to each point estimate.

\begin{table}[H]
\centering
\caption{Localized H\'ajek IPCW estimates by predicted-risk tertile at $\alpha=0.1$. CI denotes a cellwise nested-bootstrap 95\% confidence interval; UCB is the simultaneous 95\% upper confidence bound across 15 cells. Patients, events, effective $n$, and LPB are means across configuration--fold--split cells.}
\label{tab:supp-risk-tertile-hajek}
\scriptsize
\setlength{\tabcolsep}{2.3pt}
\begin{tabular}{llrrrrcl}
\toprule
Cohort & Risk & Patients & Events & Eff. $n$ & \shortstack{H\'ajek estimate\\[1pt] [95\% CI]} & UCB & \shortstack{Median LPB\\days} \\
\midrule
KIRC & Low  & 17.48 & 2.99 & 16.77 & 0.283 [0.218, 0.357] & 0.373 & 1932 \\
     & Mid  & 16.31 & 4.17 & 15.16 & 0.530 [0.471, 0.588] & 0.620 & 1173 \\
     & High & 17.41 & 9.84 & 17.06 & 0.800 [0.766, 0.832] & 0.889 & 315 \\
KIRP & Low  & 9.03  & 0.66 & 8.98  & 0.251 [0.216, 0.288] & 0.340 & 1356 \\
     & Mid  & 8.46  & 0.99 & 8.40  & 0.371 [0.319, 0.427] & 0.461 & 1252 \\
     & High & 9.31  & 2.35 & 8.71  & 0.572 [0.520, 0.624] & 0.662 & 685 \\
LUAD & Low  & 15.26 & 3.65 & 14.50 & 0.355 [0.281, 0.430] & 0.445 & 1037 \\
     & Mid  & 14.56 & 4.85 & 13.80 & 0.583 [0.493, 0.670] & 0.672 & 636 \\
     & High & 15.98 & 7.70 & 15.40 & 0.764 [0.707, 0.814] & 0.853 & 323 \\
STAD & Low  & 12.29 & 3.95 & 11.45 & 0.375 [0.307, 0.445] & 0.465 & 711 \\
     & Mid  & 11.14 & 4.75 & 10.63 & 0.626 [0.559, 0.689] & 0.715 & 386 \\
     & High & 12.37 & 5.50 & 12.08 & 0.779 [0.734, 0.821] & 0.869 & 202 \\
UCEC & Low  & 16.97 & 1.45 & 16.65 & 0.344 [0.306, 0.386] & 0.434 & 1690 \\
     & Mid  & 16.44 & 2.47 & 15.89 & 0.416 [0.358, 0.479] & 0.505 & 1293 \\
     & High & 16.79 & 4.09 & 15.85 & 0.647 [0.578, 0.713] & 0.737 & 587 \\
\bottomrule
\end{tabular}
\end{table}

All 15 simultaneous upper bounds are below 0.90, jointly placing the localized estimates below nominal under this H\'ajek estimand and hierarchy. Low risk is lowest in every cohort. KIRP had only 0.66, 0.99, and 2.35 mean events in the low-, mid-, and high-risk tertiles, respectively, illustrating how sparse subgroup event support limits conditional assessment. One KIRP mid-risk cell (MambaMIL, fold 2, split 1) contained no evaluation patients, so its ratio was undefined; the corresponding configuration--fold unit averages the other two splits. The other 14 cohort--tertile summaries each contain 270 defined ratios. We use the normalized H\'ajek ratio because an unnormalized inverse-weighted total may exceed one and is not a coverage probability. H\'ajek normalization supplies a bounded diagnostic, not a conditional guarantee or a correction for marginal Kaplan--Meier misspecification, subgroup exchangeability, or unstable censoring weights.

\subsection{Model configurations}

The configurations assess model breadth under a shared protocol. Coordinate-dependent methods receive WSI coordinates, while all configurations share the same frozen features, adaptor, hazard head, survival loss, and stopping rule. HilbertSort, the Mamba2MIL adaptation, and ABMamba are study-specific implementations informed by the cited architectures. The AEM-regularized survival adaptation adds negative attention entropy with cosine weight annealing; it is not a verbatim reproduction of the classification implementation. ABMIL and gated attention are separate configurations but share one row because both follow Ilse et al.; the 17 displayed rows therefore represent 18 configurations.

\begin{table}[t]
\centering
\caption{Configurations in the 450-run internal sweep, with architectural sources and shared-protocol specifications.}
\label{tab:supp-configurations}
\scriptsize
\setlength{\tabcolsep}{3pt}
\begin{tabular}{p{0.18\textwidth}p{0.13\textwidth}p{0.21\textwidth}p{0.38\textwidth}}
\toprule
Configuration & Family & Source & Shared-protocol specification \\
\midrule
ABMIL; gated attention & Attention & \citet{Ilse2018AttentionMIL} & Shared survival objective \\
CLAM-SB & Attention & \citet{Lu2021CLAM} & Native auxiliary loss omitted \\
DSMIL & Attention & \citet{Li2021DSMIL} & Native auxiliary loss omitted \\
DTFD & Attention & \citet{Zhang2022DTFDMIL} & Native auxiliary loss omitted \\
PTCMIL & Token clustering & \citet{Zhao2025PTCMIL} & Native auxiliary loss omitted \\
ILRA & Attention & \citet{Xiang2023ILRAMIL} & Native auxiliary loss omitted \\
DGR & Attention & \citet{Zhu2024DGRMIL} & Native auxiliary loss omitted \\
ACMIL & Attention & \citet{Zhang2024ACMIL} & Native auxiliary loss omitted \\
AEM survival adaptation & Attention & \citet{Zhang2025AEM} & Survival NLL + entropy regularizer; cosine weight annealing \\
PatchGCN & Graph/spatial & \citet{Chen2021PatchGCN} & Requires WSI coordinates \\
CAMIL & Graph/spatial & \citet{Fourkioti2024CAMIL} & Requires WSI coordinates \\
TransMIL & Transformer & \citet{Shao2021TransMIL} & Shared survival objective \\
MambaMIL & State space & \citet{Yang2024MambaMIL} & Shared survival objective \\
2D MambaMIL & State space & \citet{Zhang2025TwoDMamba} & Requires WSI coordinates \\
HilbertSort & State space & \citet{MambaBack2026} & Custom Hilbert-order serialization inspired by MambaBack \\
Mamba2MIL & State space & \citet{Zhang2024Mamba2MIL} & Adapted to the shared survival objective; CUDA/Triton \\
ABMamba & State space & \citet{Zhang2026SSRMam2MIL} & Adapted from SSR-Mam2MIL \\
\bottomrule
\end{tabular}
\end{table}

\end{document}